\documentclass[twocolumn,prl,superscriptaddress,notitlepage]{revtex4-1}

\usepackage{graphicx} 
\usepackage{booktabs}
\usepackage{siunitx}
\usepackage{url}
\usepackage{adjustbox}
\usepackage{float}

\newcolumntype{d}{S[
    input-open-uncertainty=,
    input-close-uncertainty=,
    parse-numbers = false,
    table-align-text-pre=false,
    table-align-text-post=false
 ]}

\newcommand{\djw}[1]{{\color{black} #1}}

\begin{document}
\title{Causally estimating the effect of YouTube's recommender system \\ using counterfactual bots}

\author{Homa Hosseinmardi}
\affiliation{University of Pennsylvania, Philadelphia, PA 19104}
\author{Amir Ghasemian} 
\affiliation{Yale University, New Haven, CT 06511}
\author{Miguel Rivera-Lanas}
\affiliation{University of Pennsylvania, Philadelphia, PA 19104}
\author{Manoel Horta Ribeiro}
\affiliation{EPFL, 1015 Lausanne, Switzerland}
\author{Robert West}
\affiliation{EPFL, 1015 Lausanne, Switzerland}
\author{Duncan J. Watts}
\affiliation{University of Pennsylvania, Philadelphia, PA 19104}

\keywords{Algorithm audits $|$ Experiment design $|$ Recommender systems $|$ Online extremism} 

\begin{abstract}
In recent years, critics of online platforms have raised concerns about the ability of recommendation algorithms to amplify problematic content, with potentially radicalizing consequences. However, attempts to evaluate the effect of recommenders have suffered from a lack of appropriate counterfactuals---what a user would have viewed in the absence of algorithmic recommendations---and hence cannot disentangle the effects of the algorithm from a user's intentions. 
Here we propose a method that we call ``counterfactual bots'' to causally estimate the role of algorithmic recommendations on the consumption of highly partisan content. By comparing bots that replicate real users' consumption patterns with ``counterfactual'' bots that follow rule-based trajectories, we show that, on average, relying exclusively on the recommender results in less partisan consumption, where the effect is most pronounced for heavy partisan consumers. 
Following a similar method, we also show that if partisan consumers switch to moderate content, YouTube's sidebar recommender ``forgets'' their partisan preference within roughly 30 videos regardless of their prior history, while homepage recommendations shift more gradually towards moderate content. 
Overall, our findings indicate that, at least \djw{since the algorithm changes that YouTube implemented in 2019}, individual consumption patterns mostly reflect individual preferences, where algorithmic recommendations play, if anything, a moderating role. 
\end{abstract}

\maketitle

With over 250 million active users in the US and over 2.6 billion worldwide, YouTube is among the world's largest and most engaging social media platforms. 
Moreover, while news and other related content account for a relatively small share of both production and consumption, the sheer scale of the platform means that YouTube is also one of the largest online sources of political information for Americans, roughly equivalent to Twitter~\cite{emarketer_youtube,konitzer2020measuring,businessofapps_twitter}. 
Finally, while on-platform news consumption is dominated by mainstream and moderate sources~\cite{hosseinmardi2021examining}, a relatively small but still substantial population of YouTube users consume concerning amounts of ideologically extreme~\cite{brown2022echo}, conspiratorial~\cite{hussein2020measuring}, and inflammatory content~\cite{ribeiro2020auditing}. 
The ready availability of problematic content, along with the pervasive presence of algorithmically generated recommendations on the site, has led to prominent speculation that YouTube is actively radicalizing its users via its recommender system~\cite{tufekci2018youtube,youtuberadical}. 
As has been pointed out~\cite{Arvind_YouTube_recommendation,ribeiro2023amplification,d2020fairness, sinha2016deconvolving,garimella2021political}, however, the content that users consume is some unobserved combination of their own preferences and the platform design, including the recommender, each of which influences the other in a complex feedback loop with potentially emergent properties. 
Careful empirical work is therefore needed to estimate the effect of platform design on user consumption in a way that accounts for user preferences. 

To date, empirical studies using different methodological approaches have reached somewhat different conclusions regarding the relative importance of algorithmic recommendations. While no studies find support for the alarming claims of radicalization that characterized early, anecdotal accounts, audit studies in which bots~\cite{haroon2022youtube} or humans~\cite{brown2022echo} follow rule-based viewing patterns---and platform recommendations are systematically recorded---have found that blindly following the recommender system results in ideologically biased recommendations, implying that the recommender is at least partly responsible. 
In contrast, panel studies~\cite{hosseinmardi2021examining,chen2022subscriptions} based on real user traces over many months show that the consumption of ``radical'' content on YouTube does not increase over time or with session length (on average), and is highly correlated with off-platform consumption, suggesting that user preferences are more to blame than biased recommendations~\cite{munger2022right}.

\begin{figure*}[ht]
\centering
  \includegraphics[width=0.85\textwidth]{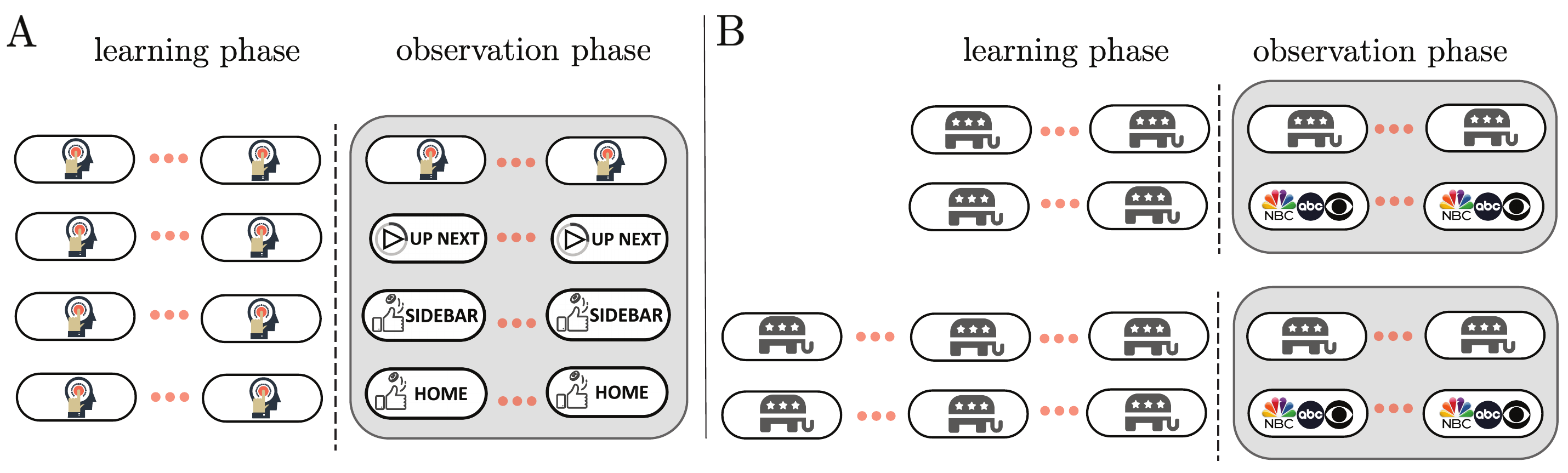}
\caption{\textbf{Overview of the counterfactual bot method to disentangle YouTube’s recommender system from user preferences utilizing counterfactual bots.} Each panel shows the trajectories (one per row) that bots traverse within the corresponding experiment. By measuring the difference in the partisanship of watched videos by control bot ($y^\mathrm{expr.}$) and watched videos by rule-base counterfactual bots ($y^\mathrm{alg.}$), our design eliminates the ``preference'' or ``choice'' component ($\hat{y}^\mathrm{pref.}= y^\mathrm{expr.}-y^\mathrm{alg.}$) of observed consumption, allowing us to estimate the causal effect of algorithmic recommendations. 
\textbf{(A)~Estimating bias of the recommender:}
Four bots watch the same history in the learning phase, whereas in the observation phase, the control bot continues to follow the real user's historical trajectory and the ``counterfactual'' bots follow simple algorithmic rules: ``up next'' (choosing the top-ranked video from the sidebar), ``random sidebar'' (choosing a random video from the sidebar), and ``random home'' (choosing a random video from the homepage).
\textbf{(B)~Estimating ``forgetting time'' of the recommender:}
Two bots start at the same time, watching the same trajectory in the learning period. The control bot will continue watching from the same trajectory in the observation phase, while the counterfactual bot will switch to watching videos of moderate leaning. To estimate the effects of different-length histories, half the bots have ``short'' (30 video) histories prior to switching, while the other half have ``long'' (120 video) histories. 
}
\label{fig:schematic_design}
\end{figure*}

Critically, neither type of study is sufficient to resolve the key causal question: how much bias do recommenders cause? 
By design, panel studies only observe what users actually clicked on, not what was recommended to them. 
As a result, they cannot rule out that the platform is recommending more extreme content than is visible in the consumption patterns, nor can they reveal what a user would have watched in the absence of recommendations.
Audit studies, meanwhile, also cannot estimate the causal effect of the recommender on biased consumption. 
 Say, for example, that a hypothetical user who ignored all recommendations ended up consuming content that is at least as biased as an otherwise identical user who only clicked on recommended content. 
 In that case, one would not conclude that the algorithm itself is biased even if the ``algorithmic'' user also consumed biased content: only if the latter were more biased than the former would the recommender be responsible for the residual bias. Just as with panel studies, audit studies do not create counterfactual comparisons of this sort and hence cannot identify the cause of the observed bias. 
 A second, related shortcoming of audit studies is that the causal (i.e., counterfactual) effect of the recommender likely depends on the type of user; specifically, how much moderate vs. extreme content they would have consumed even in the absence of recommendations. 
 Here, audit studies struggle to find the right balance between capturing rare and highly unrepresentative users who are unlikely to show up in surveys~\cite{brown2022echo} while also not assuming far higher concentrations of extreme content than is consumed by any real user~\cite{haroon2022youtube}.

In this paper, we propose a novel experimental approach, which we call ``counterfactual bots,'' designed to causally estimate the effect of algorithmic recommendations independent of user intentions. 
The bots in question are logged-in, programmatic users, each trained on the exact historical trajectory of a real user, drawn from empirical panel data encompassing $15$ months (Oct 2021-Dec 2022) of desktop browsing behavior by $87,988$ users. 

Each experiment proceeds in two phases. First, during an initial ``learning'' phase, all bots follow the same sequence of videos, ensuring that they present indistinguishable ``preferences'' to YouTube's recommender system. However, in a second ``observation'' phase, each bot is assigned to one of two types of treatment: the ``user'' treatment, in which the bot continues to follow the historical trajectory of the focal user; or a ``counterfactual'' treatment in which they follow some other rule such as clicking on the top-ranked sidebar (i.e., ``up next'') video or imitating a different type of user,~Fig.~\ref{fig:schematic_design}. 
Upon completion of each experiment, we use the YouTube API to retrieve metadata associated with each video ID in our collection, which we use to estimate the partisanship of the content (see Methods and Materials for details).
By measuring the difference in the partisanship of watched and recommended videos between user and counterfactual treatments in the observation phase, our approach eliminates the ``preference'' or ``choice'' component of observed consumption, allowing us to estimate the causal effect of algorithmic recommendations. 
An additional advantage of our design is that by training our bots on historical user data, our results have high ecological validity, meaning that they describe the effects of recommendations on real users rather than hypothetical ones. 
Finally, leveraging a large, representative historical panel allows us to estimate the effect of the recommender for different types of users---in particular, users who consume the largest amounts of problematic content. 
As noted above, these users are rare and hence are unlikely to volunteer for online experiments or surveys; however, by oversampling the ``tail'' of the distribution, we can obtain accurate estimates even for rare cases~\cite{chen2022subscriptions,yang2021online}. 


\begin{figure}[tb!]
\centering
  \includegraphics[width=0.48\textwidth]{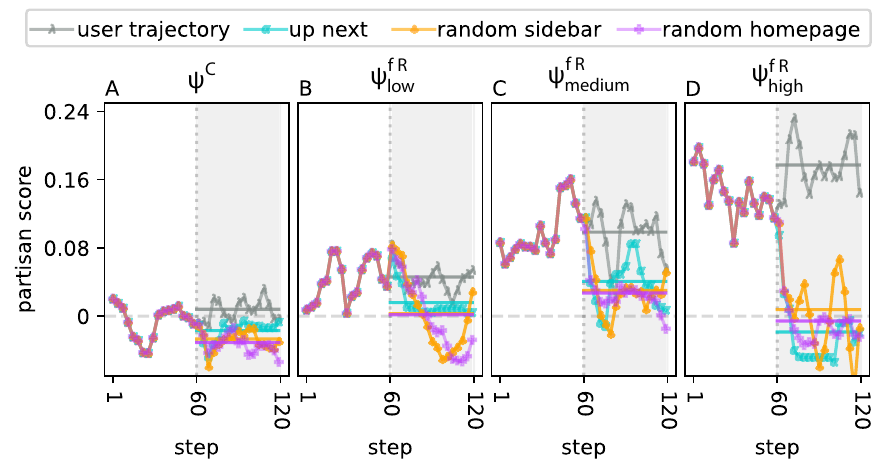}
\caption{Examples of traversed trajectories for four focal users with different news consumption archetypes from \emph{center} ($\psi^{C}$) and \emph{far-right} ($\psi^{fR}$) in the counterfactual experiment, Fig. \ref{fig:schematic_design}A. The first half is dedicated to learning  (all four bots watch the same videos at each step) and the second half (shaded grey area) is the observation phase (each of the four bots follows a separate rule). 
The y-axis provides the partisanship of watched videos at each step. The dashed line shows zero partisanship.
Solid lines show the average partisan score of all 60 watched videos in the observation phase for each path. 
}
\label{fig:examples}
\end{figure}



\begin{figure}[tb!]
\centering
  \includegraphics[width=0.48\textwidth]{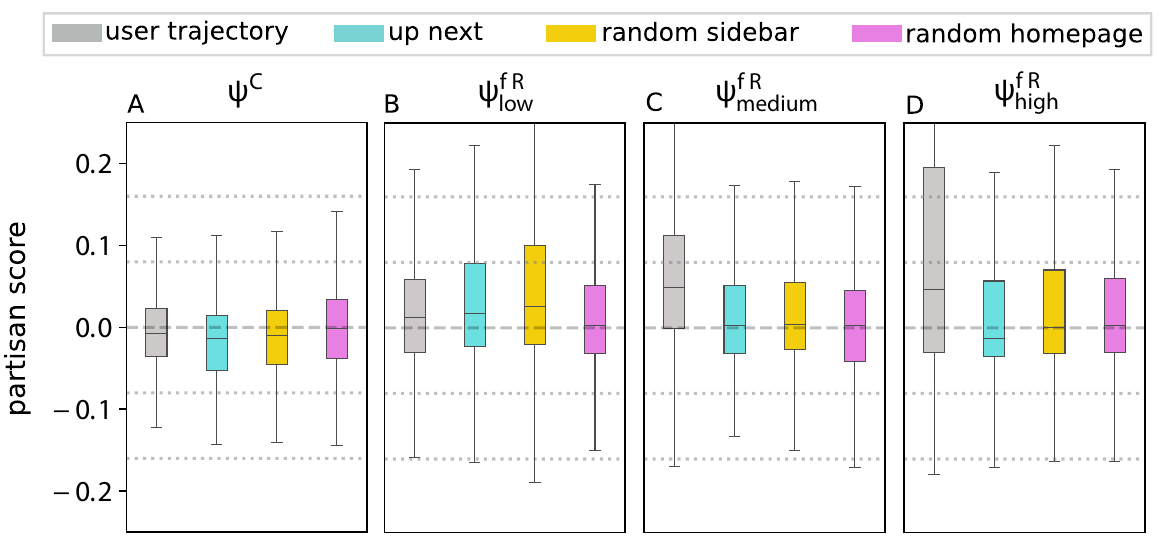}
\caption{Partisan score of the 60 watched videos by the control and counterfactual bots during the observation phase for focal users with different news consumption archetypes from \emph{center} ($\psi^{C}$) and \emph{far right} ($\psi^{fR}$) groups. Each box-plot shows the median, interquartile range, and full range of the average partisanship (the y-axis range is limited to [-0.24,0.24] for better visualization). The dashed line shows zero partisanship, and the dotted lines represent one and two standard deviations away from the mean (zero) of partisan scores.}
\label{fig:boxplots}
\end{figure}


Our analysis yields four main findings. 
First, we find that algorithmic bots, on average, receive less partisan recommendations and consume less partisan content than the corresponding ``real'' user---a result that is stronger for heavier consumers of partisan content. 
Second, we find that real users who consume ``bursts'' of highly partisan videos subsequently consume more partisan content than identical bots who subsequently follow algorithmic viewing rules. 
Third, we find that when a user switches their diet from one dominated by \emph{far-right} news content to one dominated by moderate news content, recommendations of \emph{far-right} content essentially disappear from the sidebar within 30 videos, but linger for longer in homepage recommendations. Fourth, we show that longer histories of prior \emph{far-right} consumption results in longer ``forgetting'' times of homepage recommendations, but has no impact on the forgetting time of sidebar recommendations. 
Together, our results show that platform recommendations serve, on average, to moderate a user's experience relative to following their own exogenous preferences, where the moderating effect is predominately experienced by extreme users. 
\djw{Noting that in 2019 YouTube made substantial changes to their recommendation algorithm that may have impacted the likelihood of recommending highly partisan content\cite{chen2022subscriptions}, our results suggest that at least in the post-2019 era, a user's preferences are the primary determinant of their experience.}

 \section*{Results}
Our four findings derive from two main experiments, each of which leverages counterfactual bots in somewhat different ways. 
In the first main experiment, shown schematically in Fig.~\ref{fig:schematic_design}A, the bots simulate a user who switches from replicating the behavior of a real user (during the learning phase) to one who follows a simple heuristic (during the observation phase) such as clicking on the top-ranked (aka ``up next'') video on the right side of the screen. Leveraging this design, we extract our first two main results: one estimating the causal effect of the recommender for different types of users; and one estimating the causal effect of users consuming ``bursts'' of far-right videos.    
In the second main experiment, shown schematically in Fig.~\ref{fig:schematic_design}B, the bots simulate a user ``switching'' from one set of preferences (dominated by \emph{far-right} consumption) to another (moderate consumption) and measure the ``forgetting time'' of the recommender. As with the first main experiment, we leverage the design to extract two findings: one estimating the forgetting time for a user with a ``short'' (30 video) history of far-right consumption; and one comparing the forgetting times of short and ``long'' (120 video) history consumers.

Both experiments leverage the same sample of $4,583$ users who  watched at least $140$ YouTube videos during Oct 2021-Dec 2022, drawn from a much larger ($N=87,988$) US representative desktop panel (see Methods and Materials). From this sample, we then further sampled trajectories with a length of exactly $120$ videos from each of these users by choosing a random start point between 1 and $M_i-120$, where $M_i$ is the total number of video views for the $i^{th}$ user, and taking the next 120 videos. 
The number of sampled trajectories from each user is proportional to the user's lifetime in the panel, resulting in $24,871$ unique user histories (see Methods and Materials). 
 \djw{We use channel labels provided by Ref.~\cite{boesinger2023tube2vec} and assign all videos produced by a given channel the same partisan score.} 
Next, we clustered these histories into eight news consumption ``archetypes'' $\psi^X$ ranging from $\psi^{fL}$, characterized by mostly \emph{far-left} with some centrist content, to $\psi^{fR}$, characterized by mostly \emph{far-right} content. Recognizing that within the $\psi^{fR}$ archetype there remains considerable heterogeneity regarding the relative consumption of \emph{fR} vs. other content as well as the total volume of \emph{fR} videos, we further decompose $\psi^{fR}$ into $\psi^{fR}_{\rm{low}}$, $\psi^{fR}_{\rm{medium}}$, and $\psi^{fR}_{\rm{high}}$.

\paragraph*{Estimating bias of the recommender}
In this experiment, we sampled randomly 32 histories from $\psi^{C}$ (characterized almost exclusively by centrist consumption) and ran a stratified sampling from $\psi^{fR}$, choosing 35 random histories from the $\psi^{fR}_{\rm{low}}$ group and taking all 41 and 17 histories from each of the $\psi^{fR}_{\rm{medium}}$ and $\psi_{\rm{high}}^{fR}$ ones respectively, yielding a final sample of 125 ``focal'' users.
We note that $\psi^{C}$ accounted for roughly 66\% of all histories in our sample, whereas $\psi^{fR}$ accounted for only 1.12\%; thus, our final sample over-represents heavy consumers of \emph{far-right} content, who otherwise would not appear in sufficient numbers to power our analysis.  

As noted above, the experiment comprised two phases. In the first half, the ``learning'' phase, four logged-in bots simultaneously and independently followed the trajectory of the focal user for the first $N_{\rm{learning}}=60$ videos of the focal user history. In this way, the recommender system had ample time to learn the preferences of each of the bots, but because all bots had the exact same history, they all presented the same preferences. 
In the second half, the ``observational'' phase, one of the bots (control bot) continued to watch videos from the trajectory of the same user for an additional $N_{\rm{observation}}=60$ videos, while the other three bots (counterfactual bots) switched to following one of the following rule-based trajectories: ``up next,'' in which the bot deterministically selects the first video from the sidebar recommendations; ``random sidebar,'' in which the bot randomly selects one of the top 30 videos listed in the sidebar recommendations; and ``random homepage,'' in which the bot randomly selects a video from the top 15 videos listed in the homepage recommendations. 
For each of the selected focal users, we conducted three replications of this experiment, where each replication began with identical initial conditions but varied depending on the stochastic responses of YouTube's recommender system (i.e., if two hypothetical users created the exact same profile and watched the exact same sequence of videos, their recommendations would still not be identical). 
In total, our experiment comprised four bots per replication with an average of $2.61$ replications per focal user for $125$ focal users, yielding $1,304$ independent trajectories of $120$ videos each and an estimated cumulative watch time of over $640,975$ minutes.

Fig.~\ref{fig:examples} shows four instances of the experiment for one sample focal user from each of the $\psi^{C}$, $\psi^{fR}_{\rm{low}}$, $\psi^{fR}_{\rm{medium}}$ and $\psi^{fR}_{\rm{high}}$ archetypes. As expected, the average partisanship of the videos consumed during the observation period increases with the partisanship of the focal user: whereas the bots replicating the $\psi^{C}$ users generally consume videos that fluctuate around a partisan score of $0$ (Fig.~\ref{fig:examples}A), the bots replicating the $\psi^{fR}_{\rm{low}}$, $\psi^{fR}_{\rm{medium}}$, and $\psi^{fR}_{\rm{high}}$ users consume progressively more partisan content (Fig.~\ref{fig:examples}B, C, and D respectively).  
Also as expected, the trajectories of all four bots are indistinguishable during the learning period, reflecting that they are all viewing the same sequence of videos. 
In the observation period, however, the bot trajectories diverge: whereas the control bot (grey line) continues on a similar path to the learning phase, the three counterfactual bots---up next (blue line), random sidebar (yellow line), and random homepage (purple line)---take somewhat different paths, both from the control and from each other. Notably, all three counterfactual bots are trend toward less partisan content than the control, where the gap is small in the case of the $\psi^{C}$ user (Fig.~\ref{fig:examples}A) but becomes increasingly pronounced as the partisanship of the focal user increases. In the case of the $\psi^{fR}_{\rm{high}}$ user (Fig.~\ref{fig:examples}D) the difference is highly pronounced and suggests that for extremely partisan users, the recommender actively promotes more moderate content than what the user would otherwise consume. 

\begin{table}
\centering
\
\begin{tabular}[t]{lccc}
\toprule
  & up next & random sidebar & random homepage\\ \hline
\midrule
preference  & \num{0.029}  *** & \num{0.016}  *** & \num{0.047} ***\\
($\alpha$) & {}[\num{0.027}, \num{0.032}] & {}[\num{0.013}, \num{0.019}] & {}[\num{0.045}, \num{0.050}]\\ \midrule \hline
depth  & \num{0.000} *** & \num{0.000} + & \num{0.000} +\\
($\beta_1$)& {}[\num{0.000}, \num{0.000}] & {}[\num{0.000}, \num{0.000}] & {}[\num{0.000}, \num{0.000}]\\ \midrule \hline
$n^{\rm{learning}}_{C}$ & \num{-0.001} *** & \num{0.001} *** & \num{-0.001} ***\\
 & {}[\num{-0.002}, \num{-0.001}] & {}[\num{0.001}, \num{0.001}] & {}[\num{-0.001}, \num{-0.001}]\\ \midrule \hline
$n^{\rm{learning}}_{R}$ & \num{0.000} & \num{0.001} *** & \num{0.000}\\
 & {}[\num{0.000}, \num{0.001}] & {}[\num{0.000}, \num{0.001}] & {}[\num{0.000}, \num{0.000}]\\ \midrule \hline
$n^{\rm{learning}}_{fR}$ & \num{0.003} *** & \num{0.003} *** & \num{0.002} ***\\
 & {}[\num{0.002}, \num{0.003}] & {}[\num{0.003}, \num{0.004}] & {}[\num{0.002}, \num{0.003}]\\ \hline
\midrule
$R^2$ & \num{0.055} & \num{0.034} & \num{0.066}\\
\bottomrule
\multicolumn{4}{l}{\rule{0pt}{1em}$+\, p < 0.1; *\, p < 0.05; **\, p < 0.01; ***\, p < 0.001$}\\
\end{tabular}
\caption{We define user preference as the difference in partisan score between the trajectory that the user has traversed (control bot) and the rule-based trajectory (counterfactual bot), which follows the recommendation only. Preference is positive for all three types of recommendations (up next, random sidebar, and random homepage). A higher number of \emph{C} videos in the learning phase results in a smaller difference between control and counterfactual bots, while a higher number of \emph{fR} videos has the opposite effect.}
\label{tab:trajectory_effect}
\end{table}


\begin{figure}[tb!]
\centering
  \includegraphics[width=0.5\textwidth]{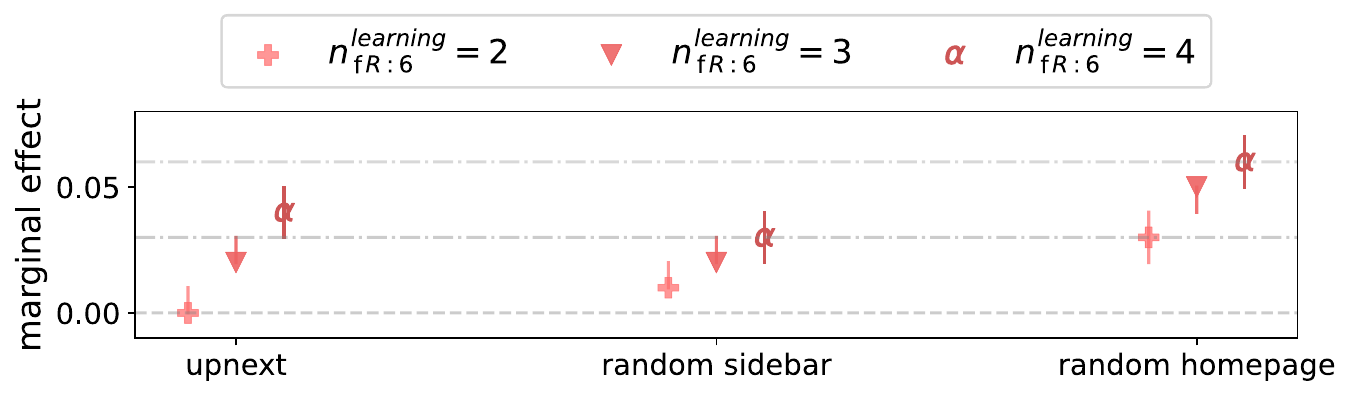}
  \label{fig:hist}
\caption{Marginal effect of bursty viewership of partisan videos (calculated using ggeffects R-package) on the user preference role in future consumption. Preference increases with higher bursts of partisan consumption. }
\label{fig:bursty}
\end{figure}

Fig.~\ref{fig:boxplots} shows these differences more systematically: each box-plot shows the median, interquartile range, and full range of the average partisanship for the watched videos by each of the four bots during the observation phase.  
Fig.~\ref{fig:boxplots}A reveals that for $\psi^{C}$ users, both counterfactual bots and control bots received relatively non-partisan recommendations, on average, and the differences between control and counterfactual bot experiences were small with effect size less than $0.2$.
Figs.~\ref{fig:boxplots}B-D show that as the partisanship of the focal users increases, the gap between partisanship of control and counterfactual bots increases, suggesting that the net effect of the recommender was, if anything, to moderate the partisanship of the user experience.  
To validate this intuition, we first compute user ``preference'' $\hat{y}^\mathrm{pref.}= y^\mathrm{expr.}-y^\mathrm{alg.}$ as the gap between the partisanship of the control bot trajectory $y^\mathrm{expr.}$ (user ``experience'' on the platform) and the partisanship of a counterfactual bot $y^\mathrm{alg.}$ (``algorithmic'' or rule-based path); thus, a positive value of $\hat{y}^\mathrm{pref.}$ corresponds to an intrinsic preference for partisan content relative to what the recommender system is recommending.

Then we regress $\hat{y}^\mathrm{pref.}_t = \alpha + \beta_1 t + \beta_2 n^{\rm{learning}}_{C}+ \beta_2 n^{\rm{learning}}_{R} + \beta_2 n^{\rm{learning}}_{fR}$ on historical features of the learning phase, including the step $t$ at which the video was watched, and the number of center videos $n^{\rm{learning}}_{C}$, the number of \emph{R} videos $n^{\rm{learning}}_{R}$, and the number of \emph{fR} videos $n^{\rm{learning}}_{fR}$ in the learning phase. 
Table~\ref{tab:trajectory_effect} shows that $\alpha$ is positive for all three types of recommendations (up next, random sidebar, and random homepage), confirming that recommendations have moderating effects relative to the focal users' intrinsic preferences. 
Furthermore, a higher number of \emph{C} videos in the learning phase ($n^{\rm{learning}}_{C}$) results in a smaller difference between control and counterfactual bots, while a higher number of \emph{fR} videos ($n^{\rm{learning}}_{fR}$) has the opposite effect. Table~\ref{tab:trajectory_effect} also shows that $\beta_1$, the coefficient for the number of steps into the observed trajectory, is not significantly different from 0, suggesting that trajectories do not become more or less extreme over time. Finally, $\alpha$ is larger for random homepage than up next, which in turn is larger than for random sidebar, suggesting that homepage recommendations are more moderate than sidebar, but that the top-ranked sidebar recommendation is more moderate than the rest of the sidebar.

To examine the robustness and generalizability of these findings, we also conducted three supplemental analyses. First, we re-analyzed the data from our experiment replacing partisan score with (a) an ``establishment'' score that captured the extent to which channel owners position themselves as non-partisan ``anti-establishment'' figures; and (b) a popularity score based on views, likes, and comments. For the establishment score, we found similar results to partisanship whereas for popularity we found no consistent effect of the recommender in either direction. Second, to check that our findings generalize to other parts of the ideological spectrum, we conducted an additional experiment for consumers of predominantly ``far-left'' partisan content, finding similar results to Figs.~\ref{fig:boxplots}. Third, to check the effect of channel subscriptions, we conducted another experiment in which the 17 $fR$-high focal users also subscribed to their three most visited channels, again finding very similar results to Fig.~\ref{fig:boxplots}.

\emph{Bursty viewership effect}.
Even if the recommender moderates a user's experience on average, it may be the case that it overreacts to ``bursts'' of partisan consumption, defined as viewership of highly partisan videos in near succession. Previous work~\cite{hosseinmardi2021examining} has found that bursts of this sort (for lengths 2, 3, and 4) predict subsequent higher consumption of partisan content but could not determine if the cause was endogenous user preferences or the exogenous response of the recommender. 
Here we revisit this question by exploiting the presence of real users in our data who consumed bursts of up to 6 videos from one of $\{C,R,fR\}$ categories during the last six videos of the learning phase. We then regress $ \hat{y}^\mathrm{pref.} = \alpha + \beta_1 n^{\rm{learning}}_{C:6} + \beta_2 n^{\rm{learning}}_{R:6}+ \beta_3 n^{\rm{learning}}_{fR:6}$, where \emph{j}:6 represents the number of videos from category $j\in\{C,R,fR\}$ in the burst. Fig.~\ref{fig:bursty} shows that the marginal prediction of preference increases for $n^{\rm{learning}}_{fR:6}\in\{2,3,4\}$ and all three types of recommendation and is positive except for $n^{\rm{learning}}_{fR:6}=2$ for up next recommendations, which is not distinguishable from $0$. Similar to our main analysis, therefore, recommendations following bursts of highly-partisan consumption offer greater moderating effects than for non-bursty consumption. 
Put another way, bursts of partisan consumption predict future consumption because they signal a change in user preferences toward more extreme content, not because the recommender is suddenly recommending more such content.


\begin{figure}[tb!]
\centering
  \includegraphics[width=0.475\textwidth]{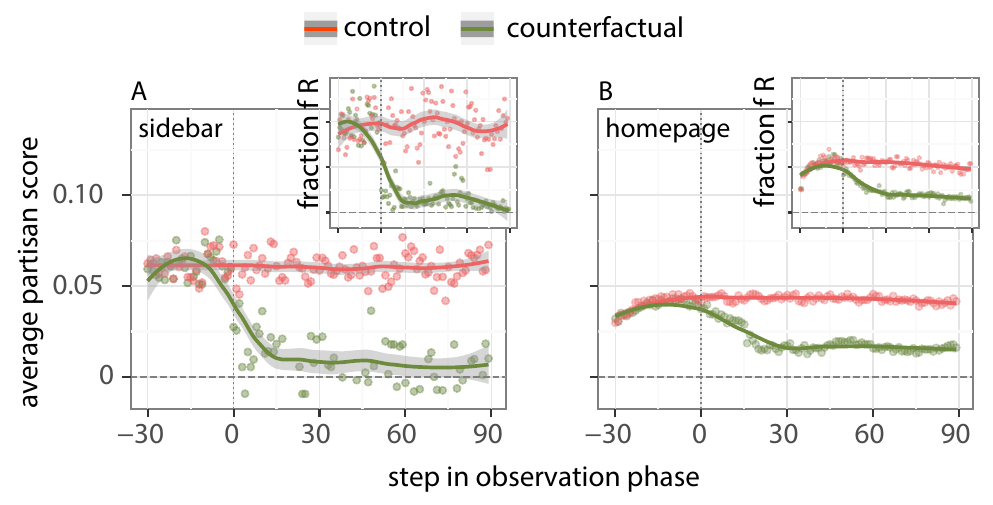}
\caption{``Forgetting'' time: A comparison of the average partisan score and the fraction of recommended \emph{fR} videos for two different paths. 
The control bot watches a 120 videos from a \emph{fR} focal user, while the counterfactual bot after watching the same 30-video history as in the control bot, transitions to videos from a \emph{center} focal user spanning 90 videos. Sidebar response to this change in consumption is immediate and partisan score converge to zero, while on the homepage side, although the average partisan score converges to moderate range, even after 90 steps, the average fraction of \emph{fR} videos remains nonzero. For better visualization, y-axis range across all panels is the same.}
\label{fig:resilience_intervention}
\end{figure}

\begin{figure}[tb!]
\centering
  \includegraphics[width=0.47\textwidth]{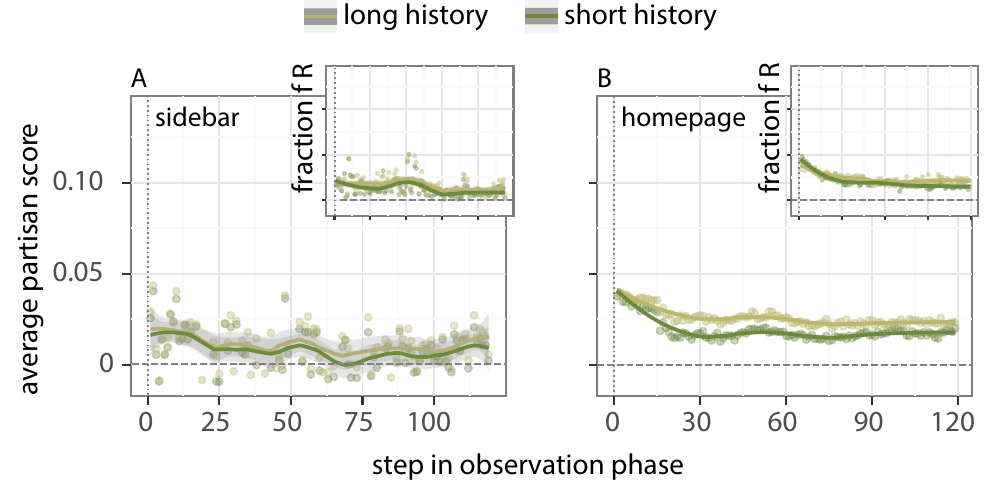}
\caption{Effect of history length: A comparison of the average partisan score and the fraction of recommended \emph{far-right} videos for two different paths with varying histories of \emph{fR} viewership. In the control arm, a bot watches a 30-video \emph{fR} history followed by a 120-video \emph{center} history, while in the treatment arm, the bot is exposed to an additional \emph{fR} history lasting 120 videos. On the sidebar, both longer and shorter history exhibit the same drop rate in terms of average partisan score and average fraction of \emph{fR} content. However, on the homepage side, longer history reduces the drop rate of partisanship on both metrics. Even after 90 steps, the average partisan score of bots with longer history remains higher than that of the shorter path. For better visualization, y-axis range across all panels is the same.}
\label{fig:resilience_intervention_history}
\end{figure}

\paragraph*{Estimating ``forgetting time'' of the recommender} Recommendation algorithms have been criticized for continuing to recommend problematic content to previously interested users long after they have lost interest in it themselves~\cite{youtuberadical}. 
To understand to which extent this is the case, we again train the bots on the trajectory of a user from the \emph{far right} end of the political spectrum, where half the bots (``short history'') imitate the user for $30$ videos and the other half (``long history'') do so for $120$ videos. 
In the second phase, both sets of bots switch to the trajectory of a different user, whose consumption is dominated by moderate and mainstream sources, and follow this user for another $120$ videos. Throughout both phases, we tracked the recommended items in the sidebar and homepage at each step and measured the progress of the average partisanship of recommended videos. In this way, we measured the rate at which the recommender ``forgets'' the prior preferences of the focal user for users with different length histories. 
We conducted the experiment for $44$ focal users---$17$ drawn from $\psi^{fR}_{\rm{high}}$ and $27$ from $\psi^{fR}_{\rm{medium}}$---where in each case the counterfactual bot was supplied by a randomly selected history from $\psi^{C}$. Replicating the experiment for each focal user three times yielded a total of $233$ trajectories comprising $45,435$ watched videos and an estimated watch time of $170,381$ minutes. We leveraged this setup to simulate two related experiments, which used data from the same underlying design in different ways.

First, we assigned a ``short-history'' bot to be the counterfactual bot and a ``long-history'' bot as the control. For both bots, therefore, the learning phase comprised $N_{\rm{learning}}=30$ and the observational phase comprised $N_{\rm{observation}}=90$, where the control watched $120$ videos from $\psi^{fR}$ group while the counterfactual bot watched $30$ videos and then switched to watching $90$ videos from a moderate content, $\psi^{C}$ (Fig.~\ref{fig:schematic_design}B). 
Fig.~\ref{fig:resilience_intervention} shows the average partisanship of sidebar and homepage recommendations for control (red line) and counterfactual (green line) bots. For sidebar recommendations (Fig.~\ref{fig:resilience_intervention}A), the counterfactual bot experiences a large and rapid decrease in partisanship relative to the control bot: within roughly 30 videos, sidebar recommendations had become indistinguishable from those recommended to a $\psi^{C}$ user, whereas those for the control bot remained almost as partisan as during the learning phase. Homepage recommendations (Fig.~\ref{fig:resilience_intervention}B), meanwhile, also decrease in partisanship for the counterfactual bot relative to the control, but tend to be less sensitive to user behavior than the sidebar: they are less partisan to begin with but also adjust less rapidly to any changes, taking roughly 30 videos to become neutral on average.  
Fig. \ref{fig:resilience_intervention}A and B insets show a similar pattern holds for the fraction of \emph{fR} videos displayed: on average, \emph{fR} videos disappear from the sidebar recommendations between 30 and 40 into the observation phase; however, a small but non-zero fraction of \emph{fR} videos continues to appear on the homepage until the end of the 90-step observation phase.

\emph{Effect of history-length in forgetting time.} To examine whether the forgetting time of the recommender depends on the length of the learning phase, we now assign the short-history bot to the control condition with $N_{\rm{learning}}=30$ and the long-history bot to the counterfactual condition with $N_{\rm{learning}}=120$, where both bots then have $N_{\rm{observation}}=120$ (Fig.~\ref{fig:schematic_design}C). Thus, the control bot in this experiment watches a total of $150$ videos ($30$ from $\psi^{fR}$ followed by $120$ from $\psi^{C}$) while the counterfactual bot watches a total of $240$ ($120$ from $\psi^{fR}$ followed by $120$ from $\psi^{C}$).  If a longer history of viewing a \emph{fR} history causes the recommender to ``remember'' the user's preference for longer, we ought to see a slower decrease in partisanship during the observation phase for the counterfactual than for the control bot. In contrast, Fig.~\ref{fig:resilience_intervention_history}A shows no such effect in the case of sidebar recommendations: although the path with a longer viewing history exhibits slightly higher average partisanship scores, both paths exhibit a similar drop rate, and both paths converge towards zero recommendations of \emph{fR} videos (Fig.~\ref{fig:resilience_intervention_history}A inset). On the other hand, Fig.~\ref{fig:resilience_intervention_history}B shows that homepage recommendations do display a slower drop rate: whereas the average partisanship of videos in both paths stabilizes around step $30$, the path with a longer history exhibits a gradual decrease that persists until the end of the observation window. The fraction of \emph{fR} videos drops along the trajectory, where from step 70 they diverge slightly (Fig.~\ref{fig:resilience_intervention_history}B Inset).

\section*{Discussion}
Online platforms like YouTube are regularly accused of amplifying politically extreme content via their recommender systems and thereby driving their users toward conspiratorial beliefs. 
Superficially, these accusations appear plausible: many users rely on recommendations to find new content; some of that content is indeed objectionable, and some users do indeed consume it. It is important to note, however, that even strong correlational evidence of this sort does not constitute evidence that the recommender itself is causing the problematic behavior. Users of online platforms also exhibit considerable agency and might have consumed the same content, or worse, even in the absence of recommendations. 

Previous empirical work has struggled to tease out the specific causal role of platform recommendations in large part because of the absence of a proper counterfactual. In some cases~\cite{hosseinmardi2021examining}, we can observe the real users' consumption but not the consumption of a counterfactual user who relied exclusively on recommendations. In other cases~\cite{haroon2022youtube,brown2022echo}, the opposite applies: we observe what a synthetic rule-following user (either a bot or a human) would be exposed to, but not what a counterfactual user who only followed their own preferences would see.  Ideally, we would like to see both the real user and their rule-following counterfactual: if the latter receives more extreme recommendations than the former, that would be evidence that the recommender is amplifying extreme content; and if it receives less extreme recommendations, that would be evidence that the recommender exerts a moderating effect.

In this paper, we have implemented precisely this design using a combination of real user data and automated bots: logged-in, programmatic users capable of following arbitrary viewing patterns. In our experiments, the bots are assigned to one of two conditions: the control bot imitates the behavior of a real ``focal'' user, whereas the counterfactual bot initially imitates the behavior of the same user but then switches to a different behavior such as clicking on the top-ranked sidebar (aka ``up next'') recommendation (Fig.~\ref{fig:schematic_design}). By comparing the experience of the counterfactual bot with that of the control, we can estimate the causal effect of the recommender. Moreover, by selecting different types of focal users---defined by the amount of \emph{far-right} (\emph{fR}) vs. centrist (\emph{C}) content they consumed---we can measure how the causal effect varies with user behavior.

Our results suggest that, on average, relying solely on the recommender results in a more moderate experience on YouTube relative to the real user, where the effect is mostly driven by extreme users (Fig. \ref{fig:boxplots} and Table \ref{tab:trajectory_effect}) and for users who consume ``bursts'' of \emph{fR} videos (Fig. \ref{fig:bursty}). 
Further, we found that when consumers of partisan content change to moderate content, the sidebar reacts quickly and \emph{fR} content, on average, decreases to zero after 30 steps, while homepage recommendations react more slowly (Fig.~\ref{fig:resilience_intervention}). We also found that the ``forgetting rate'' for the homepage is longer for users with longer histories, whereas sidebar recommendations are unaffected (Fig.~\ref{fig:resilience_intervention_history}).

Overall, our study reinforces previous work~\cite{munger2022right,hosseinmardi2021examining,Guess2023ChronolOrder, ribeiro2023amplification,robertson2023users} that places individual human preferences at the center of platform dynamics. While recommendations and other platform affordances no doubt shape user experiences to some degree~\cite{gonzalez2023asymmetric}, our results suggest that popular narratives~\cite{tufekci2018youtube, youtuberadical, pariser2011filter, haidt2023social} about the widespread and profound manipulative impact of algorithms are overstated. 
This is not to say that highly problematic content does not exist on social media platforms, that it does not have harmful effects on those who consume it, or that platforms should not be held responsible for mitigating these effects. Rather, by shifting the emphasis of the concern from presumed biases in algorithms to the factors governing the supply and demand of problematic content, social media companies and their critics can more accurately target the source of the problem, which may transcend any one platform however large. For example, recent work~\cite{horta2023deplatforming} shows that shutting down the right-wing social media site Parler had little impact on the overall consumption of conspiratorial content, as users simply replaced their diets of such content via other sources on the web.

Although we believe our contribution constitutes a meaningful advance for studying the causal effects of platform design, it nonetheless has limitations. \djw{First, as noted earlier, in 2019 YouTube implemented significant changes to its algorithm that it claimed reduced watch time of ``borderline content and harmful misinformation'' by 50-70\%  \cite{chen2022subscriptions}.  It is therefore possible that to some extent the difference between our findings and pre-2019 claims of the radicalizing effects of YouTube's algorithm can be attributed to changes to the algorithm. Unfortunately, testing this hypothesis would require recreating YouTube as it existed prior to the change, which is to our knowledge impossible; thus, our findings should be interpreted as applying only to the post-2019 period.}  Second, our experiments were conducted in early 2023 whereas our empirical data was recorded between October 2021 through Dec 2022. Although we are not aware of any major changes to YouTube's moderation policy or recommendation system in the intervening months, and we conducted multiple iterations of each experiment in order to account for randomness and other time-varying factors, our experiment was not a true field experiment. 
\djw{Third, our empirical panel data is restricted to desktop users and hence does not include YouTube consumption on mobile devices, which could potentially be different. }
Fourth, for feasibility, we sped up the bot viewing to simulate several months of real user activity in two to three days. Although we do not believe that speeding up the watch time meaningfully altered the recommender's reactions, we cannot rule that the same experiment conducted over many months would yield different results. \djw{Fifth, the scoring is done at the channel level, which is not entirely accurate as there may be differences in partisanship levels across videos within a channel. Future work would benefit from a video-level scoring approach to identify partisanship of content more precisely.}
In spite of these limitations, we hope our work will stimulate researchers of socio-technical systems to adopt counterfactual bot designs. We believe these designs strike a useful balance between taking real user behavior seriously and exploiting the flexibility, speed, and data-recording capabilities of programmatic users. In this sense, our study can also be viewed as a proof of concept for an approach to studying the interactions between humans and algorithms across many online platforms and services, not just YouTube.

\section*{Methods and Materials}

\subsection*{Dataset}
Our data are derived from Nielsen's nationally representative desktop web panel, which tracks individuals' visits to URLs from October 2021 to December 2022, including a total of $87,988$ panelists. Each YouTube video has a unique identifier embedded in its URL. 
By parsing the recorded URLs, we find the subset of $48,026$ users who have at least one recorded YouTube video viewership. 
To post a video on YouTube, a user must create a channel with a unique name and channel ID. For all unique video IDs collected from Nielsen and recorded in the experiments, we used the YouTube API to retrieve the corresponding channel ID, as well as metadata such as the video's category, title, and duration. We then use the channel IDs to assign a partisanship score to each video based on the political leaning of its channel. Table \ref{basic_stats} provides more details on data statistics. 

\subsection*{``User history'' selection}

Our unit of analysis in this paper is ``user history,'' where we focused on heavy consumers of \emph{far-right} content. 
To ensure a comprehensive and representative selection of user histories, we employed a systematic approach. Initially, we searched across all $4,583$ users who had watched a minimum of $140$ YouTube videos and sampled trajectories with a length of $120$ videos by choosing a random start point between 1 and $M_i-120$, where $M_i$ is the total number of YouTube video views for the $i^{th}$ user. From each user, we randomly selected multiple histories according to their lifetime on the panel, resulting in $24,871$ histories, with $12,969$ having at least 1 minute of news consumption (from $3,089$ unique news users). 
We continued by grouping histories based on their news consumption archetype using the first $N_{\rm{learining}}=60$ videos. We did so to avoid looking into future consumption, which will be used for evaluation purposes (in this way, when a user history is assigned to an archetype, the observation period is not known, and there is no leakage of future information in the assignment of users to experiments, i.e., we do not keep users who already have a high consumption of \emph{fR} in the observation period). 
Following the same approach as Ref.~\cite{hosseinmardi2021examining}, we characterized every user history in terms of their normalized news viewership vector. We adopted a source-based approach where we assigned all videos produced by a channel the same partisan score. To derive the political partisanship scores of channels, we leveraged the embeddings of approximately $7.5$ million channels provided by Ref.~\cite{boesinger2023tube2vec}, which incorporated the Reddit embeddings developed by Waller and Anderson \cite{waller_quantifying_2021}. 
The scores were validated using existing lists of left- and right-wing YouTube channels (e.g., \cite{hosseinmardi2021examining}), resulting in a rank correlation of $0.65$. Further, in a crowdsourcing task, the authors found agreement between embedding and crowd workers to be above $80\%$, indicating the robustness of their approach. 
Overall, $20\%$ of the collected video IDs don't have a partisan score attached to them, and for the presented results in the main text, we have dropped such videos from our analysis. To validate the robustness of these findings, we have replicated our analysis where missing values are imputed. With the average partisan score zero and the standard deviation $\sigma=0.08$, any video with a partisan score in range $(-\sigma,~\sigma)$ is labeled as \emph{C}; $[~-\sigma,~-2 \sigma)$ and $[\sigma,2\sigma)$ are labeled as \emph{L} and \emph{R}, respectively; anything to the left of left $[-\sigma,-2\sigma)$ is labeled as \emph{far-left}, \emph{fL}, and anything to the right of right is labeled as \emph{far right}, \emph{fR}. The normalized viewership vector of $i$th history is $\nu_i$, whose $j$th entry $\nu_{ij}$ corresponds to the fraction of viewership of $i$th user-history from $j$th category ($j \in \{{\emph{fL,~L,~C,~R,~fR}}\}$). 
We then used hierarchical clustering to assign each user history to one of $K=8$ communities of similar YouTube news diets. We ended up with $144$ histories with heavy \emph{fR} consumption (from $90$ unique YouTube users), which we used to select histories for this study. To better understand the underlying patterns within this category of behavior, again, we employed a hierarchical clustering algorithm, grouping \emph{fR} histories into three distinct archetypes, each representing a unique pattern of consumption of \emph{fR} videos. To ensure a balanced and representative analysis of the results, we either select all histories or randomly select a subset, depending on the size of the cluster. In all subsequent analyses, we weighted these samples to accurately reflect the true distribution of histories within the overall population.

\begin{table}[tbhp!]
\centering
\caption{Data descriptive statistics.}
\begin{tabular}{lr}
\hline
Number of panelists & $87,988$  \\ \hline
Number of YouTube consumers & $48,026$  \\ \hline
Total number of watched trajectories & $1,537$  \\ \hline
Number of watched videos by bots & $201,915$\\ \hline
Estimated total watched time (minutes) & $811,356$ \\ \hline
\end{tabular}\label{basic_stats}
\end{table}


\subsection*{Designing bots}

Overall, we created more than 150 Google accounts for this study, and each account has been used across multiple experiments. Our test experiments show that logged-out browsing behaves differently from logged-in accounts regarding the similarity of recommended items with the watched history. 
Further, as personalization in the logged-out approach is via browser cookies, it is specific to a particular browser session. It imposes technical limitations for very long sessions, which may take days, as any interruption and browser reset may lead to loss of historical information. Therefore, we run all experiments with logged-in bots. The web crawler includes functionality designed to reset YouTube accounts to a clean state. This feature enables the crawler to log into the user's account, access the ``Your data in YouTube'' section, and clear the watch history. By doing so, all user activity data on YouTube is effectively removed, and the recommendations are reset to a state similar to that of an incognito window, based on our knowledge. The empirical validity of this approach has been confirmed by the experiments, where initial measurements do not indicate any presence of previously watched topics on this account. 

In all experiments, each trajectory is divided into two parts: learning and observation. In the learning phase, bots are first ``primed'' with real user histories by watching their videos from the corresponding focal user. This is equivalent to creating multiple copies of the same account with personalized recommendations. Recognizing the importance of variations in the ``Watch Time'' for the recommendation system to learn users' interests in different topics \cite{covington2016deep,YouTube_recommendation}, we allocate a watch time to each video that is proportionate (half of the actual watch time) to the real user's video viewing duration from Nielsen data. Moreover, we introduce pauses (half of the actual pause duration) between videos that correspond to the behavior of real users, thereby enhancing the accuracy of mimicking their viewing patterns. For both watch time and idle times, we set a maximum limit of $10$ minutes and $20$ minutes, respectively, to ensure the feasibility of the experiments. Upon completing each experiment, we retrieve metadata associated with each video ID in our collection using the YouTube API. Only a small percentage, less than 3.1\%, of video IDs do not produce metadata from the API.

\begin{acknowledgements}
\textbf{Acknowledgments}: We are grateful to the Nielsen Company for access to their desktop panel data, and to B. Sissenich, S. Sherman, H. Baberwal, and E. Grimaldi for ongoing support.
Additionally, H.H., M.R., and D.J.W. are grateful for the financial support provided by Richard Jay Mack and the Carnegie Corporation of New York (Grant G-F-20-57741). 
A. G. is supported by the National Science Foundation under Grant No.~2030859 to the Computing Research Association for the CIFellows Project.
M.H.R. and R.W. acknowledge support from the Swiss National Science Foundation (grant 200021\_185043) and the European Union (TAILOR, grant 952215).
\end{acknowledgements}


%


\begin{thebibliography}{28}%
\makeatletter
\providecommand \@ifxundefined [1]{%
 \@ifx{#1\undefined}
}%
\providecommand \@ifnum [1]{%
 \ifnum #1\expandafter \@firstoftwo
 \else \expandafter \@secondoftwo
 \fi
}%
\providecommand \@ifx [1]{%
 \ifx #1\expandafter \@firstoftwo
 \else \expandafter \@secondoftwo
 \fi
}%
\providecommand \natexlab [1]{#1}%
\providecommand \enquote  [1]{``#1''}%
\providecommand \bibnamefont  [1]{#1}%
\providecommand \bibfnamefont [1]{#1}%
\providecommand \citenamefont [1]{#1}%
\providecommand \href@noop [0]{\@secondoftwo}%
\providecommand \href [0]{\begingroup \@sanitize@url \@href}%
\providecommand \@href[1]{\@@startlink{#1}\@@href}%
\providecommand \@@href[1]{\endgroup#1\@@endlink}%
\providecommand \@sanitize@url [0]{\catcode `\\12\catcode `\$12\catcode
  `\&12\catcode `\#12\catcode `\^12\catcode `\_12\catcode `\%12\relax}%
\providecommand \@@startlink[1]{}%
\providecommand \@@endlink[0]{}%
\providecommand \url  [0]{\begingroup\@sanitize@url \@url }%
\providecommand \@url [1]{\endgroup\@href {#1}{\urlprefix }}%
\providecommand \urlprefix  [0]{URL }%
\providecommand \Eprint [0]{\href }%
\providecommand \doibase [0]{http://dx.doi.org/}%
\providecommand \selectlanguage [0]{\@gobble}%
\providecommand \bibinfo  [0]{\@secondoftwo}%
\providecommand \bibfield  [0]{\@secondoftwo}%
\providecommand \translation [1]{[#1]}%
\providecommand \BibitemOpen [0]{}%
\providecommand \bibitemStop [0]{}%
\providecommand \bibitemNoStop [0]{.\EOS\space}%
\providecommand \EOS [0]{\spacefactor3000\relax}%
\providecommand \BibitemShut  [1]{\csname bibitem#1\endcsname}%
\let\auto@bib@innerbib\@empty
\bibitem [{\citenamefont {{A Schomer}}(2020)}]{emarketer_youtube}%
  \BibitemOpen
  \bibfield  {author} {\bibinfo {author} {\bibnamefont {{A Schomer}}},\ }\href
  {\url{https://www.emarketer.com/content/us-youtube-advertising-2020}}
  {\enquote {\bibinfo {title} {{US} {YouTube} advertising 2020},}\ } (\bibinfo
  {year} {2020})\BibitemShut {NoStop}%
\bibitem [{\citenamefont {Konitzer}\ \emph {et~al.}(2020)\citenamefont
  {Konitzer}, \citenamefont {Allen}, \citenamefont {Eckman}, \citenamefont
  {Howland}, \citenamefont {Mobius}, \citenamefont {Rothschild},\ and\
  \citenamefont {Watts}}]{konitzer2020measuring}%
  \BibitemOpen
  \bibfield  {author} {\bibinfo {author} {\bibfnamefont {T.}~\bibnamefont
  {Konitzer}}, \bibinfo {author} {\bibfnamefont {J.}~\bibnamefont {Allen}},
  \bibinfo {author} {\bibfnamefont {S.}~\bibnamefont {Eckman}}, \bibinfo
  {author} {\bibfnamefont {B.}~\bibnamefont {Howland}}, \bibinfo {author}
  {\bibfnamefont {M.~M.}\ \bibnamefont {Mobius}}, \bibinfo {author}
  {\bibfnamefont {D.~M.}\ \bibnamefont {Rothschild}}, \ and\ \bibinfo {author}
  {\bibfnamefont {D.}~\bibnamefont {Watts}},\ }\href@noop {} {\bibfield
  {journal} {\bibinfo  {journal} {Available at SSRN 3548690}\ } (\bibinfo
  {year} {2020})}\BibitemShut {NoStop}%
\bibitem [{\citenamefont {{M Iqbal}}(2021)}]{businessofapps_twitter}%
  \BibitemOpen
  \bibfield  {author} {\bibinfo {author} {\bibnamefont {{M Iqbal}}},\ }\href
  {\url{https://www.businessofapps.com/data/twitter-statistics/}} {\enquote
  {\bibinfo {title} {{Twitter} revenue and usage statistics 2020},}\ }
  (\bibinfo {year} {2021})\BibitemShut {NoStop}%
\bibitem [{\citenamefont {Hosseinmardi}\ \emph {et~al.}(2021)\citenamefont
  {Hosseinmardi}, \citenamefont {Ghasemian}, \citenamefont {Clauset},
  \citenamefont {Mobius}, \citenamefont {Rothschild},\ and\ \citenamefont
  {Watts}}]{hosseinmardi2021examining}%
  \BibitemOpen
  \bibfield  {author} {\bibinfo {author} {\bibfnamefont {H.}~\bibnamefont
  {Hosseinmardi}}, \bibinfo {author} {\bibfnamefont {A.}~\bibnamefont
  {Ghasemian}}, \bibinfo {author} {\bibfnamefont {A.}~\bibnamefont {Clauset}},
  \bibinfo {author} {\bibfnamefont {M.}~\bibnamefont {Mobius}}, \bibinfo
  {author} {\bibfnamefont {D.~M.}\ \bibnamefont {Rothschild}}, \ and\ \bibinfo
  {author} {\bibfnamefont {D.~J.}\ \bibnamefont {Watts}},\ }\href@noop {}
  {\bibfield  {journal} {\bibinfo  {journal} {Proceedings of the National
  Academy of Sciences}\ }\textbf {\bibinfo {volume} {118}},\ \bibinfo {pages}
  {e2101967118} (\bibinfo {year} {2021})}\BibitemShut {NoStop}%
\bibitem [{\citenamefont {Brown}\ \emph {et~al.}(2022)\citenamefont {Brown},
  \citenamefont {Bisbee}, \citenamefont {Lai}, \citenamefont {Bonneau},
  \citenamefont {Nagler},\ and\ \citenamefont {Tucker}}]{brown2022echo}%
  \BibitemOpen
  \bibfield  {author} {\bibinfo {author} {\bibfnamefont {M.~A.}\ \bibnamefont
  {Brown}}, \bibinfo {author} {\bibfnamefont {J.}~\bibnamefont {Bisbee}},
  \bibinfo {author} {\bibfnamefont {A.}~\bibnamefont {Lai}}, \bibinfo {author}
  {\bibfnamefont {R.}~\bibnamefont {Bonneau}}, \bibinfo {author} {\bibfnamefont
  {J.}~\bibnamefont {Nagler}}, \ and\ \bibinfo {author} {\bibfnamefont {J.~A.}\
  \bibnamefont {Tucker}},\ }\href@noop {} {\bibfield  {journal} {\bibinfo
  {journal} {Available at SSRN 4114905}\ } (\bibinfo {year}
  {2022})}\BibitemShut {NoStop}%
\bibitem [{\citenamefont {Hussein}\ \emph {et~al.}(2020)\citenamefont
  {Hussein}, \citenamefont {Juneja},\ and\ \citenamefont
  {Mitra}}]{hussein2020measuring}%
  \BibitemOpen
  \bibfield  {author} {\bibinfo {author} {\bibfnamefont {E.}~\bibnamefont
  {Hussein}}, \bibinfo {author} {\bibfnamefont {P.}~\bibnamefont {Juneja}}, \
  and\ \bibinfo {author} {\bibfnamefont {T.}~\bibnamefont {Mitra}},\
  }\href@noop {} {\bibfield  {journal} {\bibinfo  {journal} {Proceedings of the
  ACM on Human-Computer Interaction}\ }\textbf {\bibinfo {volume} {4}},\
  \bibinfo {pages} {1} (\bibinfo {year} {2020})}\BibitemShut {NoStop}%
\bibitem [{\citenamefont {Ribeiro}\ \emph {et~al.}(2020)\citenamefont
  {Ribeiro}, \citenamefont {Ottoni}, \citenamefont {West}, \citenamefont
  {Almeida},\ and\ \citenamefont {Meira~Jr}}]{ribeiro2020auditing}%
  \BibitemOpen
  \bibfield  {author} {\bibinfo {author} {\bibfnamefont {M.~H.}\ \bibnamefont
  {Ribeiro}}, \bibinfo {author} {\bibfnamefont {R.}~\bibnamefont {Ottoni}},
  \bibinfo {author} {\bibfnamefont {R.}~\bibnamefont {West}}, \bibinfo {author}
  {\bibfnamefont {V.~A.}\ \bibnamefont {Almeida}}, \ and\ \bibinfo {author}
  {\bibfnamefont {W.}~\bibnamefont {Meira~Jr}},\ }in\ \href@noop {} {\emph
  {\bibinfo {booktitle} {Proceedings of the 2020 conference on fairness,
  accountability, and transparency}}}\ (\bibinfo {year} {2020})\ pp.\ \bibinfo
  {pages} {131--141}\BibitemShut {NoStop}%
\bibitem [{\citenamefont {Tufekci}(2018)}]{tufekci2018youtube}%
  \BibitemOpen
  \bibfield  {author} {\bibinfo {author} {\bibfnamefont {Z.}~\bibnamefont
  {Tufekci}},\ }\href
  {https://www.nytimes.com/2018/03/10/opinion/sunday/youtube-politics-radical.html}
  {\enquote {\bibinfo {title} {Youtube, the great radicalizer},}\ } (\bibinfo
  {year} {2018})\BibitemShut {NoStop}%
\bibitem [{\citenamefont {Roose}(2019)}]{youtuberadical}%
  \BibitemOpen
  \bibfield  {author} {\bibinfo {author} {\bibfnamefont {K.}~\bibnamefont
  {Roose}},\ }\href
  {https://www.nytimes.com/interactive/2019/06/08/technology/youtube-radical.html}
  {\enquote {\bibinfo {title} {The making of a {YouTube} radical},}\ }
  (\bibinfo {year} {2019})\BibitemShut {NoStop}%
\bibitem [{\citenamefont {Narayanan}(2023)}]{Arvind_YouTube_recommendation}%
  \BibitemOpen
  \bibfield  {author} {\bibinfo {author} {\bibfnamefont {A.}~\bibnamefont
  {Narayanan}},\ }\href
  {https://knightcolumbia.org/content/understanding-social-media-recommendation-algorithms}
  {\enquote {\bibinfo {title} {Understanding social media recommendation
  algorithms},}\ } (\bibinfo {year} {2023})\BibitemShut {NoStop}%
\bibitem [{\citenamefont {Ribeiro~Horta}\ \emph {et~al.}(2023)\citenamefont
  {Ribeiro~Horta}, \citenamefont {Veselovsky},\ and\ \citenamefont
  {West}}]{ribeiro2023amplification}%
  \BibitemOpen
  \bibfield  {author} {\bibinfo {author} {\bibfnamefont {M.}~\bibnamefont
  {Ribeiro~Horta}}, \bibinfo {author} {\bibfnamefont {V.}~\bibnamefont
  {Veselovsky}}, \ and\ \bibinfo {author} {\bibfnamefont {R.}~\bibnamefont
  {West}},\ }\href@noop {} {\bibfield  {journal} {\bibinfo  {journal} {preprint
  arXiv:2302.11225}\ } (\bibinfo {year} {2023})}\BibitemShut {NoStop}%
\bibitem [{\citenamefont {D'Amour}\ \emph {et~al.}(2020)\citenamefont
  {D'Amour}, \citenamefont {Srinivasan}, \citenamefont {Atwood}, \citenamefont
  {Baljekar}, \citenamefont {Sculley},\ and\ \citenamefont
  {Halpern}}]{d2020fairness}%
  \BibitemOpen
  \bibfield  {author} {\bibinfo {author} {\bibfnamefont {A.}~\bibnamefont
  {D'Amour}}, \bibinfo {author} {\bibfnamefont {H.}~\bibnamefont {Srinivasan}},
  \bibinfo {author} {\bibfnamefont {J.}~\bibnamefont {Atwood}}, \bibinfo
  {author} {\bibfnamefont {P.}~\bibnamefont {Baljekar}}, \bibinfo {author}
  {\bibfnamefont {D.}~\bibnamefont {Sculley}}, \ and\ \bibinfo {author}
  {\bibfnamefont {Y.}~\bibnamefont {Halpern}},\ }in\ \href@noop {} {\emph
  {\bibinfo {booktitle} {Proceedings of the 2020 Conference on Fairness,
  Accountability, and Transparency}}}\ (\bibinfo {year} {2020})\ pp.\ \bibinfo
  {pages} {525--534}\BibitemShut {NoStop}%
\bibitem [{\citenamefont {Sinha}\ \emph {et~al.}(2016)\citenamefont {Sinha},
  \citenamefont {Gleich},\ and\ \citenamefont
  {Ramani}}]{sinha2016deconvolving}%
  \BibitemOpen
  \bibfield  {author} {\bibinfo {author} {\bibfnamefont {A.}~\bibnamefont
  {Sinha}}, \bibinfo {author} {\bibfnamefont {D.~F.}\ \bibnamefont {Gleich}}, \
  and\ \bibinfo {author} {\bibfnamefont {K.}~\bibnamefont {Ramani}},\
  }\href@noop {} {\bibfield  {journal} {\bibinfo  {journal} {Advances in neural
  information processing systems}\ }\textbf {\bibinfo {volume} {29}} (\bibinfo
  {year} {2016})}\BibitemShut {NoStop}%
\bibitem [{\citenamefont {Garimella}\ \emph {et~al.}(2021)\citenamefont
  {Garimella}, \citenamefont {Smith}, \citenamefont {Weiss},\ and\
  \citenamefont {West}}]{garimella2021political}%
  \BibitemOpen
  \bibfield  {author} {\bibinfo {author} {\bibfnamefont {K.}~\bibnamefont
  {Garimella}}, \bibinfo {author} {\bibfnamefont {T.}~\bibnamefont {Smith}},
  \bibinfo {author} {\bibfnamefont {R.}~\bibnamefont {Weiss}}, \ and\ \bibinfo
  {author} {\bibfnamefont {R.}~\bibnamefont {West}},\ }in\ \href@noop {} {\emph
  {\bibinfo {booktitle} {Proceedings of the International AAAI Conference on
  Web and Social Media}}},\ Vol.~\bibinfo {volume} {15}\ (\bibinfo {year}
  {2021})\ pp.\ \bibinfo {pages} {152--162}\BibitemShut {NoStop}%
\bibitem [{\citenamefont {Haroon}\ \emph {et~al.}(2022)\citenamefont {Haroon},
  \citenamefont {Chhabra}, \citenamefont {Liu}, \citenamefont {Mohapatra},
  \citenamefont {Shafiq},\ and\ \citenamefont
  {Wojcieszak}}]{haroon2022youtube}%
  \BibitemOpen
  \bibfield  {author} {\bibinfo {author} {\bibfnamefont {M.}~\bibnamefont
  {Haroon}}, \bibinfo {author} {\bibfnamefont {A.}~\bibnamefont {Chhabra}},
  \bibinfo {author} {\bibfnamefont {X.}~\bibnamefont {Liu}}, \bibinfo {author}
  {\bibfnamefont {P.}~\bibnamefont {Mohapatra}}, \bibinfo {author}
  {\bibfnamefont {Z.}~\bibnamefont {Shafiq}}, \ and\ \bibinfo {author}
  {\bibfnamefont {M.}~\bibnamefont {Wojcieszak}},\ }\href@noop {} {\bibfield
  {journal} {\bibinfo  {journal} {preprint arXiv:2203.10666}\ } (\bibinfo
  {year} {2022})}\BibitemShut {NoStop}%
\bibitem [{\citenamefont {Chen}\ \emph {et~al.}(2022)\citenamefont {Chen},
  \citenamefont {Nyhan}, \citenamefont {Reifler}, \citenamefont {Robertson},\
  and\ \citenamefont {Wilson}}]{chen2022subscriptions}%
  \BibitemOpen
  \bibfield  {author} {\bibinfo {author} {\bibfnamefont {A.~Y.}\ \bibnamefont
  {Chen}}, \bibinfo {author} {\bibfnamefont {B.}~\bibnamefont {Nyhan}},
  \bibinfo {author} {\bibfnamefont {J.}~\bibnamefont {Reifler}}, \bibinfo
  {author} {\bibfnamefont {R.~E.}\ \bibnamefont {Robertson}}, \ and\ \bibinfo
  {author} {\bibfnamefont {C.}~\bibnamefont {Wilson}},\ }\href@noop {}
  {\bibfield  {journal} {\bibinfo  {journal} {preprint arXiv:2204.10921}\ }
  (\bibinfo {year} {2022})}\BibitemShut {NoStop}%
\bibitem [{\citenamefont {Munger}\ and\ \citenamefont
  {Phillips}(2022)}]{munger2022right}%
  \BibitemOpen
  \bibfield  {author} {\bibinfo {author} {\bibfnamefont {K.}~\bibnamefont
  {Munger}}\ and\ \bibinfo {author} {\bibfnamefont {J.}~\bibnamefont
  {Phillips}},\ }\href@noop {} {\bibfield  {journal} {\bibinfo  {journal} {The
  International Journal of Press/Politics}\ }\textbf {\bibinfo {volume} {27}},\
  \bibinfo {pages} {186} (\bibinfo {year} {2022})}\BibitemShut {NoStop}%
\bibitem [{\citenamefont {Yang}\ and\ \citenamefont
  {Gonz{\'a}lez-Bail{\'o}n}(2021)}]{yang2021online}%
  \BibitemOpen
  \bibfield  {author} {\bibinfo {author} {\bibfnamefont {T.}~\bibnamefont
  {Yang}}\ and\ \bibinfo {author} {\bibfnamefont {S.}~\bibnamefont
  {Gonz{\'a}lez-Bail{\'o}n}},\ }\href@noop {} {\bibfield  {journal} {\bibinfo
  {journal} {Available at SSRN 3954565}\ } (\bibinfo {year}
  {2021})}\BibitemShut {NoStop}%
\bibitem [{\citenamefont {Guess}\ \emph {et~al.}(2023)\citenamefont {Guess},
  \citenamefont {Malhotra}, \citenamefont {Pan}, \citenamefont {Barber{\'a}},
  \citenamefont {Allcott}, \citenamefont {Brown}, \citenamefont
  {Crespo-Tenorio}, \citenamefont {Dimmery}, \citenamefont {Freelon},
  \citenamefont {Gentzkow} \emph {et~al.}}]{Guess2023ChronolOrder}%
  \BibitemOpen
  \bibfield  {author} {\bibinfo {author} {\bibfnamefont {A.~M.}\ \bibnamefont
  {Guess}}, \bibinfo {author} {\bibfnamefont {N.}~\bibnamefont {Malhotra}},
  \bibinfo {author} {\bibfnamefont {J.}~\bibnamefont {Pan}}, \bibinfo {author}
  {\bibfnamefont {P.}~\bibnamefont {Barber{\'a}}}, \bibinfo {author}
  {\bibfnamefont {H.}~\bibnamefont {Allcott}}, \bibinfo {author} {\bibfnamefont
  {T.}~\bibnamefont {Brown}}, \bibinfo {author} {\bibfnamefont
  {A.}~\bibnamefont {Crespo-Tenorio}}, \bibinfo {author} {\bibfnamefont
  {D.}~\bibnamefont {Dimmery}}, \bibinfo {author} {\bibfnamefont
  {D.}~\bibnamefont {Freelon}}, \bibinfo {author} {\bibfnamefont
  {M.}~\bibnamefont {Gentzkow}},  \emph {et~al.},\ }\href@noop {} {\bibfield
  {journal} {\bibinfo  {journal} {Science}\ }\textbf {\bibinfo {volume}
  {381}},\ \bibinfo {pages} {398} (\bibinfo {year} {2023})}\BibitemShut
  {NoStop}%
\bibitem [{\citenamefont {Robertson}\ \emph {et~al.}(2023)\citenamefont
  {Robertson}, \citenamefont {Green}, \citenamefont {Ruck}, \citenamefont
  {Ognyanova}, \citenamefont {Wilson},\ and\ \citenamefont
  {Lazer}}]{robertson2023users}%
  \BibitemOpen
  \bibfield  {author} {\bibinfo {author} {\bibfnamefont {R.~E.}\ \bibnamefont
  {Robertson}}, \bibinfo {author} {\bibfnamefont {J.}~\bibnamefont {Green}},
  \bibinfo {author} {\bibfnamefont {D.~J.}\ \bibnamefont {Ruck}}, \bibinfo
  {author} {\bibfnamefont {K.}~\bibnamefont {Ognyanova}}, \bibinfo {author}
  {\bibfnamefont {C.}~\bibnamefont {Wilson}}, \ and\ \bibinfo {author}
  {\bibfnamefont {D.}~\bibnamefont {Lazer}},\ }\href@noop {} {\bibfield
  {journal} {\bibinfo  {journal} {Nature}\ ,\ \bibinfo {pages} {1}} (\bibinfo
  {year} {2023})}\BibitemShut {NoStop}%
\bibitem [{\citenamefont {Gonz{\'a}lez-Bail{\'o}n}\ \emph
  {et~al.}(2023)\citenamefont {Gonz{\'a}lez-Bail{\'o}n}, \citenamefont {Lazer},
  \citenamefont {Barber{\'a}}, \citenamefont {Zhang}, \citenamefont {Allcott},
  \citenamefont {Brown}, \citenamefont {Crespo-Tenorio}, \citenamefont
  {Freelon}, \citenamefont {Gentzkow}, \citenamefont {Guess} \emph
  {et~al.}}]{gonzalez2023asymmetric}%
  \BibitemOpen
  \bibfield  {author} {\bibinfo {author} {\bibfnamefont {S.}~\bibnamefont
  {Gonz{\'a}lez-Bail{\'o}n}}, \bibinfo {author} {\bibfnamefont
  {D.}~\bibnamefont {Lazer}}, \bibinfo {author} {\bibfnamefont
  {P.}~\bibnamefont {Barber{\'a}}}, \bibinfo {author} {\bibfnamefont
  {M.}~\bibnamefont {Zhang}}, \bibinfo {author} {\bibfnamefont
  {H.}~\bibnamefont {Allcott}}, \bibinfo {author} {\bibfnamefont
  {T.}~\bibnamefont {Brown}}, \bibinfo {author} {\bibfnamefont
  {A.}~\bibnamefont {Crespo-Tenorio}}, \bibinfo {author} {\bibfnamefont
  {D.}~\bibnamefont {Freelon}}, \bibinfo {author} {\bibfnamefont
  {M.}~\bibnamefont {Gentzkow}}, \bibinfo {author} {\bibfnamefont {A.~M.}\
  \bibnamefont {Guess}},  \emph {et~al.},\ }\href@noop {} {\bibfield  {journal}
  {\bibinfo  {journal} {Science}\ }\textbf {\bibinfo {volume} {381}},\ \bibinfo
  {pages} {392} (\bibinfo {year} {2023})}\BibitemShut {NoStop}%
\bibitem [{\citenamefont {Pariser}(2011)}]{pariser2011filter}%
  \BibitemOpen
  \bibfield  {author} {\bibinfo {author} {\bibfnamefont {E.}~\bibnamefont
  {Pariser}},\ }\href@noop {} {\emph {\bibinfo {title} {The filter bubble: How
  the new personalized web is changing what we read and how we think}}}\
  (\bibinfo {year} {2011})\BibitemShut {NoStop}%
\bibitem [{\citenamefont {Haidt}\ and\ \citenamefont
  {Twenge}(2023)}]{haidt2023social}%
  \BibitemOpen
  \bibfield  {author} {\bibinfo {author} {\bibfnamefont {J.}~\bibnamefont
  {Haidt}}\ and\ \bibinfo {author} {\bibfnamefont {J.}~\bibnamefont {Twenge}},\
  }\href@noop {} {\bibfield  {journal} {\bibinfo  {journal} {Unpublished
  manuscript, New York University.}\ } (\bibinfo {year} {2023})}\BibitemShut
  {NoStop}%
\bibitem [{\citenamefont {Horta~Ribeiro}\ \emph {et~al.}(2023)\citenamefont
  {Horta~Ribeiro}, \citenamefont {Hosseinmardi}, \citenamefont {West},\ and\
  \citenamefont {Watts}}]{horta2023deplatforming}%
  \BibitemOpen
  \bibfield  {author} {\bibinfo {author} {\bibfnamefont {M.}~\bibnamefont
  {Horta~Ribeiro}}, \bibinfo {author} {\bibfnamefont {H.}~\bibnamefont
  {Hosseinmardi}}, \bibinfo {author} {\bibfnamefont {R.}~\bibnamefont {West}},
  \ and\ \bibinfo {author} {\bibfnamefont {D.~J.}\ \bibnamefont {Watts}},\
  }\href@noop {} {\bibfield  {journal} {\bibinfo  {journal} {PNAS nexus}\
  }\textbf {\bibinfo {volume} {2}},\ \bibinfo {pages} {pgad035} (\bibinfo
  {year} {2023})}\BibitemShut {NoStop}%
\bibitem [{\citenamefont {Boesinger}\ \emph {et~al.}(2023)\citenamefont
  {Boesinger}, \citenamefont {Ribeiro}, \citenamefont {Veselovsky},\ and\
  \citenamefont {West}}]{boesinger2023tube2vec}%
  \BibitemOpen
  \bibfield  {author} {\bibinfo {author} {\bibfnamefont {L.}~\bibnamefont
  {Boesinger}}, \bibinfo {author} {\bibfnamefont {M.~H.}\ \bibnamefont
  {Ribeiro}}, \bibinfo {author} {\bibfnamefont {V.}~\bibnamefont {Veselovsky}},
  \ and\ \bibinfo {author} {\bibfnamefont {R.}~\bibnamefont {West}},\
  }\href@noop {} {\bibfield  {journal} {\bibinfo  {journal} {preprint
  arXiv:2306.17298}\ } (\bibinfo {year} {2023})}\BibitemShut {NoStop}%
\bibitem [{\citenamefont {Waller}\ and\ \citenamefont
  {Anderson}(2021)}]{waller_quantifying_2021}%
  \BibitemOpen
  \bibfield  {author} {\bibinfo {author} {\bibfnamefont {I.}~\bibnamefont
  {Waller}}\ and\ \bibinfo {author} {\bibfnamefont {A.}~\bibnamefont
  {Anderson}},\ }\href {\doibase 10.1038/s41586-021-04167-x} {\bibfield
  {journal} {\bibinfo  {journal} {Nature}\ }\textbf {\bibinfo {volume} {600}},\
  \bibinfo {pages} {264} (\bibinfo {year} {2021})},\ \bibinfo {note} {number:
  7888 Publisher: Nature Publishing Group}\BibitemShut {NoStop}%
\bibitem [{\citenamefont {Covington}\ \emph {et~al.}(2016)\citenamefont
  {Covington}, \citenamefont {Adams},\ and\ \citenamefont
  {Sargin}}]{covington2016deep}%
  \BibitemOpen
  \bibfield  {author} {\bibinfo {author} {\bibfnamefont {P.}~\bibnamefont
  {Covington}}, \bibinfo {author} {\bibfnamefont {J.}~\bibnamefont {Adams}}, \
  and\ \bibinfo {author} {\bibfnamefont {E.}~\bibnamefont {Sargin}},\ }in\
  \href@noop {} {\emph {\bibinfo {booktitle} {Proceedings of the 10th ACM
  conference on recommender systems}}}\ (\bibinfo {year} {2016})\ pp.\ \bibinfo
  {pages} {191--198}\BibitemShut {NoStop}%
\bibitem [{\citenamefont {Goodrow}(2021)}]{YouTube_recommendation}%
  \BibitemOpen
  \bibfield  {author} {\bibinfo {author} {\bibfnamefont {C.}~\bibnamefont
  {Goodrow}},\ }\href
  {https://blog.youtube/inside-youtube/on-youtubes-recommendation-system/}
  {\enquote {\bibinfo {title} {On {YouTube}’s recommendation system},}\ }
  (\bibinfo {year} {2021})\BibitemShut {NoStop}%
\end{thebibliography}

\end{document}